\documentstyle[12pt]{article}
\begin{document}
\setlength{\baselineskip}{0.30in}
\newcommand{\beq}{\begin{equation}}
\newcommand{\eeq}{\end{equation}}
 {\hbox to\hsize{Feb. 1993 \hfill UM-AC-93-02}}

\begin{center}
\vglue .06in
{\Large \bf
{New Upper Limits on the
Tau Neutrino Mass from Primordial Helium
Considerations}}\\[.5in]

{\bf A.D. Dolgov \footnote{Permanent address: ITEP 113259, Moscow,
Russia.} and I.Z.~Rothstein}
\\[.05in]
{\it{The Randall Laboratory of Physics\\
University of Michigan, Ann Arbor, MI 48109}}\\[.15in]

{Abstract}\\[-.1in]
\end{center}
\begin{quotation}
In this paper we reconsider recently derived bounds on $MeV$ tau neutrinos,
taking into account previously unaccounted for effects. We find that,
assuming that the neutrino life-time is longer than $O(100~sec)$,
the constraint $N_{eff}<3.6$
rules out $\nu_{\tau}$ masses in the range
$0.5~(MeV)<m_{\nu_\tau}<35~(MeV)$ for Majorana neutrinos and
$0.74~(MeV)<m_{\nu_\tau}<35~(MeV)$ for Dirac neutrinos.
Given that the present laboratory bound is 35 MeV, our results lower
the present bound to $0.5$ and $0.74$ for Majorana and Dirac neutrinos
respectively.
\end{quotation}

\newpage
Despite a considerable experimental effort it is still unknown whether or not
 neutrinos
have a nonzero mass. The following upper bounds
are known  \cite{pd}:
$$ m_{\nu_e} < 10 \, eV  \eqno (1a) $$
$$m_{\nu_\mu} <270 \, KeV \eqno (1b)$$
$$m_{\nu_\tau} <35 \, MeV \eqno (1c)$$
More stringent bounds can be determined from cosmological considerations.
Assuming that there is no cosmological constant, the masses of
light  ($m_\nu < O(100\,MeV)$) stable neutrinos
 are
bounded by the Gerstein-Zeldovich limit \cite{gz}
which can be written as
\beq
m_\nu < 380\, eV
\left( {10^{10} \over t_U (yrs)} -h_{100} \right) ^2 ,
\eeq
where $t_U$ is the Universe age and $h_{100}= H /100\,km/sec/Mpc$
is the dimensionless Hubble parameter. By most estimates
$t_U >12 \, Gyr$ and $0.5<h<1$. Thus the cosmological upper bound
the neutrino mass (all weakly interacting flavors)
is 40 eV. If there is a non zero cosmological
constant
the bound is somewhat less restrictive, $m_\nu<
200\, eV$ \cite{sz}.
These bounds are obtained by ensuring that the
energy density of relic cosmic neutrinos be less than the closure density,
 and  do not apply
 if neutrinos are unstable with life-times smaller than the
the age of the universe.
  Recently, it has been pointed out [4] that nucleosynthesis considerations
can further constrain the mass of an unstable  $\tau$ neutrino.

Nucleosynthesis calculations along with
data on light element abundances constrains the number of effective
neutrino species
contributing to the cosmic energy density $N_\nu$,
to be less than 3.6 \cite{sky}
(Note that in ref. [4] a slightly stronger bound $N_\nu <3.4$ was used).
This bound
is a consequence of the fact that the rate at which the universe cools
depends on the total number of species contributing to the
cosmological energy density which in turn determines the light element
abundances.
Previously, it was thought that
the contribution from a heavy neutrino species with $m> O(MeV)$
could be
neglected since its energy density was assumed to be Boltzmann suppressed.
However, after a massive neutrino decouples
and becomes non-relativistic its' energy density grows relative to that
of a massless species. Therefore, if the number density of a massive species
at freeze out is on the order of the number density of a massless species,
then the heavy species may have a greater effect on the light element abundance
than a massless species.

In this paper we reconsider the bounds derived in ref. [4] taking into
account a few effects which were neglected in that paper. Our results
for the number densities
are in reasonable agreement with theirs for the case of massive Majorana
neutrinos but differ in the case of massive Dirac neutrinos.
There is a rather large difference between
our respective results for the effective number of neutrinos species
contributed by the heavy $\nu_{\tau}$.
We find a more stringent bound which allows no window between the
experimental bound and the bounds derived here.

In the standard calculations of the relic abundance of a particle
species  disappearing
due to annihilation, the following two essential assumptions are
made: The  particles in question remain
in kinetic equilibrium, and Boltzmann statistics are applicable.
In this case the Boltzmann kinetic equation
can be reduced to an ordinary differential equation.
If the species considered
is  relativistic or non-relativistic at the time of decoupling,
the equation can be greatly simplified
and relatively
accurate analytical solutions can be found. However,
for a species which is semi-relativistic when it decouples, there is
no known analytical approximation and one has to
integrate the Boltzmann equation numerically.
As such, we will present some of the details of the
calculations.

If we are concerned with the abundance of species $\psi$
 then the Boltzmann equation is given by (we specialize to the case
$\psi+2\rightarrow 3+4)$
\beq
{dn_{\psi}\over{dt}}+3Hn_{\psi}={g\over{(2\pi)^3}}\int C[f]
{d^3p_{\psi}\over{E_{\psi}}}.
\eeq
where $H=\dot R /R$ is the Hubble parameter.
In this notation
\beq n_{\psi}=g\int {d^3p \over{(2\pi)^3}} f_{\psi}
(E,t), \eeq
$f_{\psi}
(E,t)$ is the phase space density of species $\psi$, and $g$ is the number
of degrees of freedom for $\psi$.
$C[f]$ is the collision term given by
\begin{eqnarray}
{g\over{(2\pi)^3}}\int C[f]
{d^3p_{\psi}\over{E_{\psi}}} &=& -\delta^4(\sum p)(2\pi)^4
\int d\Pi_\psi d\Pi_{2}d\Pi_{3}d\Pi_{4}  \nonumber \\
& & \left[
\mid \!M_{\psi+2\rightarrow 3+4} \! \mid^2
(
f_\psi f_{2}(1\pm f_3)(1\pm f_4))\right. \nonumber \\ & &
-\mid \!M_{3+4\rightarrow \psi+2} \! \mid^2(
f_3f_4(1\pm f_\psi )\left.
(1\pm 2))\right]],
\end{eqnarray}
and
\beq
d\Pi_i={d^3p_i\over{(2\pi)^3(2E_i)}}.
\eeq
The amplitude $M$ is summed over initial and final spins.
Assuming
that the initial and final states are in kinetic equilibrium
we may write
$f_a=n_a exp(-E/T)/n_a^{EQ}$.
Eq.(4) may be written as
\beq
{dn_{\psi}\over{dt}}+3{\dot R\over{R}}n_{\psi}=- g_\psi g_2\int
d\Pi_\psi d\Pi_2 \left[\sigma v_{Moller}exp(-(E_\psi+E_2)/T)
{(n_\psi n_2 -n_\psi^{EQ} n_2 ^{EQ})\over{n_\psi^{EQ}n_2^{EQ}}}\right]
.\eeq
Following the notation of ref \cite{gg} we have defined
\beq v_{Moller}={F\over{4E_\psi E_2}}=[\mid \!\vec{v_\psi}-\vec{v_2} \!
\mid^2-\mid \! \vec{v_\psi} \times \vec{v_2} \! \mid^2]^{1/2}, \eeq
where $F$ is the particle flux.
In the non-relativistic limit the Moller velocity reduces to the
relative velocity.

It is convenient to introduce the dimensionless variable $x=m/T$ and
the relative number density $r=n/n_0$,
 where $m$ is the mass of annihilating
particles and $n_0$ is conserved in a comoving volume\footnote{
Often the
number density $n$ is normalized to the entropy density $s$. This choice
is convenient because
$s$ is usually conserved in the comoving volume.
However, if a species is annihilating while it has considerable energy
density, then when it drops out of equilibrium the entropy will
$not$ be conserved.}.
In terms of these quantities the Boltzmann equation takes the
form:
\beq
{dr \over dx} ={n_0 \over x(\dot T /T)} \langle \sigma v_{Moller}
\rangle (r^2 -r^2_{EQ})
\eeq

It can be shown that \cite{gg}
\beq
<\sigma v_{Moller}>=\int^\infty_{4m^2} {\sigma (s-4m^2)\sqrt{s}
K_1(\sqrt{s}/T)\over{8m^4TK_2^2(m/T)}}.
\eeq
In this equation $K_1$ and $K_2$ are the modified Bessel functions
and $s=(p_\psi+p_2)^2$.
The squared amplitude integrated over the final particle phase space,
 for the annihilation of two Majorana neutrinos with mass $m_{\nu_\tau}$
into fermions with mass $m_f$ is given
by
\begin{eqnarray}
{1\over{4}}\sum_{spins}
 \int d\Pi_3 d\Pi_4 \mid \! M \! \mid^2_{M}&=&
8\pi G_F^2 {w\over{s}}\left[(C_V^2+C_A^2)(m_{\nu_\tau}^2(m_f^2-s/2))
+(C_A^2-C_V^2)m_f^2(3m^2_{\nu_\tau}-s/2)\right] \nonumber \\ & &
+{32\over{12}}\pi G_F^2{w\over{s^2}}(C_V^2+C_A^2)\left[{w^2\over{2}}
(s/2-m_{\nu_{\tau}})
+{s^2\over{4}}(s+2m_f^2)  \right]
.\end{eqnarray}
The analogous expression for the case of annihilating Dirac neutrinos
is given by
\begin{eqnarray}
 {1\over{4}}\sum_{spins}\int d\Pi_3 d\Pi_4 \mid \! M \! \mid^2_{D}&=&
32\pi G_F^2 {w\over{48}}\left[(C_V^2+C_A^2)(s-m_f^2-m_{\nu_\tau}^2w^2/s^2
\right] \nonumber \\
& &+ 4\pi G_F^2 {wm_f^2\over{s}}(C_V^2-C_A^2)\left(s/2-m_{\nu_\tau}^2
\right)
{}.
\end{eqnarray}
We have defined $w=(s^2-4sm_f)$, where $m_f$ is the mass of the final
state particle, which for our purposes is the electron, or the
other neutrinos species which we will take to be massless.
In the non-relativistic
limit the cross section for the annihilation of Dirac neutrinos
reduces to
\begin{eqnarray}
\sigma_{D} v_{rel} & = & {G_F^2\over{2\pi}}m_{\nu_\tau}^2(1-z)^{1/2}
(C_V^2+C_A^2)\left[1+{\beta^2\over{6}}(5-2z^2+{3\over{1-z^2}})\right]
\nonumber \\& &+
(C_V^2-C_A^2)\left[ {z^2\over{2}}\left(1
+\beta^2({1\over{2}}+{1\over{2(1-z^2)}})\right)
\right],
\end{eqnarray}
where $z=m_f/M$, $m_f$ is the decay product mass
 and $\beta$ is the velocity in the center of mass frame.
$v_{rel}=2\beta$ in the rest frame of the plasma.
We include this expression because our result differs slightly from those
stated previously in the literature \cite{gordy} \cite{ko}.

To perform the numerical integration it is necessary to
 specify the function
$n_0 (T)$ as well as the function $T(t)$. In the
simplest case the energy density is dominated by relativistic particles
in thermal equilibrium $\dot T =-HT$ and $n_0 \propto T^3$.
In what follows we take $n_0=0.181 T^3$ which corresponds to the equilibrium
number density of massless fermions with two helicity states. The Hubble
parameter is given by the expression
\beq
H=\sqrt {8\pi \rho /3m_{Pl}^2} = 1.66\sqrt g_* T^2/M_{Pl}
\eeq
In this simple case all the quantities are defined in terms of the
plasma temperature $T$ and the numerical
integration is straightforward.
When there is a mixture of relativistic
and nonrelativistic (or semirelativistic) species with a conserved number
of nonrelativistic particles, one must substitute for $g_*$
the expression
\beq
g_* =g_*^{Rel} (1 + \rho ^{NR} / \rho ^{Rel} )
\eeq
where $\rho^{Rel}$ and $\rho ^{NR}$ are the energy densities of relativistic
and nonrelativistic species respectively. If we neglect the energy exchange
between the relativistic and nonrelativistic species the rate of the
cooling would be the same, $\dot T =-HT$ and we have to
solve kinetic equation (8) subject to (14). For the choice of normalization
$n_0 =
0.181 T^3$  eq.(14) takes the form:
\beq
g_* =g_*^{Rel} + 0.554 \langle E \rangle  r /T
\eeq
with $\langle E \rangle $
being the  average energy per particle for the massive fermions.
The massive neutrinos will have a thermal distribution as long
as they remain in kinetic equilibrium, which is true down to temperatures
 $T_0 \approx 2$ MeV. Below this temperature
the distribution function in the phase space changes from
$\exp( {-\sqrt{ (m^2 +p^2)} /T }) $ to
$\exp( {-\sqrt {(m^2/T_0^2 + p^2 /T^2)}}) $.
This change in the distribution function will modify the Boltzmann equation.
For large masses it is not expected that
these
changes will have any significant effect.
We have calculated affect of such changes and
found them to be less the two percent for masses above 1 $MeV$.
For masses
$m<1$ MeV, the neutrinos decoupled while they were still
 relativistic, so their number density will not be suppressed relative to a
massless species.
Thus in the case of $m<1~MeV$, $r=1$, and the massive neutrinos
 distribution in phase space
is given by $\exp (-p/T)$.
Given this distribution function
the
average energy is $\langle E \rangle \approx (m^2+0.414mT+3.151T^2)^{1/2}$.
This approximate expression is accurate up to 0.5\% and was used in our
numerical work for the case of light neutrinos. We should also point out
that
for smaller masses,
$m=1-5$ MeV the Boltzmann approximation is not accurate and the helium
abundance will decrease by 5-10\% \cite{dk}. However, as we shall see, this
will not affect our bounds.

Exchange of energy between massless and massive particles gives rise to
a faster cooling. Covariant energy conservation demands that
\beq
{\dot T \over T} \approx -H\left(1+{0.2r \over {g_*^{rel}+0.42r{m\over{T}}}}
\right)
-{0.14\dot {r}({3\over 2}+{m \over T})\over{g_*^{rel}+{0.42r{m \over{T}}}}}
.\eeq
This is valid if $\rho^{NR} <\rho^{ReL}$ and the massive and massless
particles are in thermal contact. If this inequality did not hold
the cooling rate
would be different, and eq.(8) would not be valid
since it was based on the assumption that
the massive particles were in kinetic equilibrium.

We have numerically solved kinetic eq.(8) for
 $g_*^{rel} =9$  and $g_*^{nr}$ given by
eq. (15). This corresponds to the case of two light
neutrinos and one heavy.
 Note that $g_*^{nr}$  depends on the unknown function
$r(T)$ which is determined from the solution of the kinetic equation.
Our results for the freeze out number density of the massive
Majorana and Dirac tau-neutrinos are presented in fig. 1.
We then calculated the $n/p$ ratio taking into account that $r$ is not
a constant but decreases due annihilation
of the heavy neutrinos. Furthermore, we
used the exact expression for the average energy density
of the heavy neutrinos by integrating the distribution function.

Bounds on the neutrino mass are derived
by computing the net increase in helium production due to the
presence of the heavy neutrino species.
In accordance with the standard nucleosynthesis calculations
almost all neutrons which survived down to the temperature
$T_\gamma =0.065 $ MeV turn into $He^4$.
Thus, by comparing the neutron to proton ratio calculated for two massless
neutrinos and one neutrino with mass $m_{\nu_{\tau}}$, to the ratio
calculated for a variable number of massless neutrino species we may bound the
neutrino mass.
If the $n/p$ ratio yield for a given mass $m_{\nu_{\tau}}$ exceeds
the ratio calculated for 3.6 massless species, then that mass is ruled
out.
For the case of Dirac neutrinos we assume that the right handed
species will be populated if $m_D<0.74~MeV$\cite{fm}.
{}From these considerations we find that, assuming the neutrino
lifetime is longer than $O(100)~sec$, the following limits apply

\beq  0.5< m_M < 35  ~(MeV),\eeq
\beq  0.74< m_D < 35 ~(MeV). \eeq

Tau neutrinos with their mass in the region considered must be
unstable in accordance with the Gerstein-Zeldovich limit. For our bounds
to be valid the neutrino must not decay prior to
primordial nucleosynthesis. Since the characteristic temperature scale
is near 0.1 MeV the life-time should be larger than or of the
order of 100 sec.
For life-times shorter than 1 $sec$\footnote{ This is assuming that
the neutrino decays only into particles present in the standard model}
, there will be no bound for
Majorana neutrinos, but for Dirac neutrinos there will be small
region of excluded masses near 1 $MeV$.

One could
deduce bounds on both $m_{\nu_\tau}$ and on $\tau_{\nu_\tau}$
accounting for the
decay of $\nu_\tau$ in the kinetic equation governing both the number
density of $\nu_\tau$ and the $n/p$-ratio. The final $n/p$-ratio depends
not only the life-time of $\nu_\tau$ but also on the type of particles
in the final state.
A considerable effect associated
with the decay
might emerge  from the distortion of the electron neutrino spectrum
if there is a decay into $\nu_e$. This is analogous
to the distortion of the spectrum of the electron neutrinos due to
electron-positron annihilation in the standard scenario at the level about
1\% found recently \cite{df},\cite{td}.
The size of the effect in the case of massive
$\nu_\tau$ annihilation or decay should be bigger because the
(hypothetical) mass of $\nu_\tau$ is assumed to be larger than $m_e=0.5$ MeV.
We hope to take all these effects into account in the subsequent publication.
\vskip.1in
\centerline{\bf{Acknowledgements}}
The authors would like to thank Terry Walker for corroborating our results
on the neutron to proton ratio for the case of massless neutrinos, as well
as Dima Dolgov for assistance with numerical work.
The authors also benefitted from
conversation with J. Primack.

\end{document}